\documentclass{fmj2010}
\begin{document}
   \title{Hadronic $\gamma$-ray emission  from extragalactic mini radio lobes}

   \author{M. Kino\inst{1}\thanks{motoki.kino@nao.ac.jp}
          \and
          K. Asano \inst{2}
          }

   \institute{National Astronomical Observatory of Japan, Tokyo 181-8588, Japan
         \and
             Interactive Research Center for Science,
             Tokyo Institute of Technology, Tokyo 152-88550, Japan
             }

   \abstract{

Hadronic emission from
parsec size radio lobes in active galactic nuclei (AGN) is discussed.
The lobes are composed of shocked jet plasma
and expected to be filled with high energy particles.
By using the Monte Carlo simulation,
we calculate the photon spectra from the lobes
including photo-meson interaction processes.
When the synchrotron emission from primary electrons is bright,  
synchrotron-self-Compton component is dominant in $\gamma$-ray bands.
The hadronic emission from the lobes 
can be dominated in $\gamma$-ray bands
when the primary emission is not very bright.
Proton synchrotron component arises at sub MeV band.
The synchrotron emission radiated 
from secondary $e^{\pm}$ pairs produced 
via photo-meson cascade emerges in $\sim$GeV-TeV energy ranges. 
These high energy emission signatures
provide a test for proton accelerations in young AGN jets.  }

   \maketitle
%

\section{Introduction}

The progress of VLBI
(Very Long Baseline Interferometry) observations
reveal the existence of compact radio-loud 
active galactic nuclei (AGN)
with its linear size $LS<1~{\rm kpc}$
(e.g., 
Readhead et al. 1996;
O'Dea \& Baum 1997).
These compact radio sources are currently
understood as young progenitor of large radio
galaxies (e.g., Fanti 2009 for review).
Furthermore,
recent investigations discover smaller
radio lobes with $LS\sim {\cal O}(10~{\rm pc})$ 
(e.g., Snellen et al. 2004; Orienti et al. 2008). 
It is well known that recurrent radio sources  
also possess mini lobes inside them
(e.g., Walker et al. 2000).
High energy emissions from these
compact radio sources have been recently explored  
by some authors (e.g., Stawarz et al. 2008; Kino et al. 2009).
However, the previous works focus on the leptonic emissions
and little is known about a hadronic emission in them.

Since mini radio lobes with $LS\sim 10~{\rm pc}$ 
are closely located near the AGN core, 
the lobes are expected to be in dense
external radiation fields (Fig.~1).
Furthermore, together with electrons, 
the shocks accelerate protons as well.
Therefore high energy 
protons are naturally expected to be filled with the radio lobes
although its total amount is still unknown.
Therefore $p\gamma$ interaction is inevitable 
for the mini lobes. 
Here, 
we examine  $p\gamma$ interactions
between high energy protons and surrounding target photons.
As for external photons, 
we focus on UV photons from the standard accretion disks.


\section{Model}

Following the
standard picture of radio lobe  formation in AGNs 
(e.g., Begelman et al. 1984), 
here we briefly review the standard picture of radio lobes.
When a jet interacts with surrounding ambient matter,
most of its kinetic energy is dissipated via strong shocks.
The termination point at the tip of the jet 
is called as hot spots.
The hot spot is identified as the reverse
shocked region of the decelerated jet.
The shocked jet plasma leak from the hot spot and 
they form radio lobes. Therefore we can say that the
radio lobes are the remnant of the decelerated jets 
which contain relativistic particles.
In this work, we examine the case when 
the lobe contains relativistic protons.

\subsection{Proton acceleration and coolings}

Hot spots in powerful radio galaxies are one of 
the promising sights for proton acceleration 
(e.g., Rachen and Biermann 1993). 
Here we show that hot spots in mini lobes are also
plausible sights for proton acceleration.
Hereafter the maximum energy of shock accelerated
protons is denoted as $E_{p,\rm max}=\gamma_{p}m_{p}c^{2}$.
The proton acceleration time scale at the  hot spots 
$t_{p,\rm acc,hs}=\xi_{p} E_{p}/(eB_{\rm  hs}c)$ 
is estimated as
\begin{eqnarray}
t_{p,\rm acc,hs}
\approx 3.5~ 
\left(\frac{\xi_{p}}{100}\right)
\left(\frac{E_{p}}{10^{18}~{\rm eV}}\right)
\left(\frac{B_{\rm hs}}{10^{-1}~{\rm G}}\right)^{-1}~{\rm yr} ,
\end{eqnarray}
where  
$B_{\rm hs}$, and
$\xi_{p}$ are the magnetic field strength, and
the ratio of the acceleration timescale and Larmor timescale 
for protons at the hot spot, respectively.
Although there is no direct estimate for $B_{\rm hs}$,
the magnetic field strength in  mini lobes ($B$) 
have been recently estimated as
$B \sim 10-100 ~{\rm mG}$ 
by VLBI observations (Orienti et al. 2008). 
Since the magnetic field strength 
at the hot spot $B_{\rm hs}$ should be comparable or larger than  $B$.
Here we set $B_{\rm hs}\sim 100~{\rm mG}$.
The value of $\xi_{p}$ is a free parameter.
For blazars, it is constrained as $\xi_{p}> 10$ 
(Aharonian et al. 2002).
There is little constraint on  $\xi_{p}$ for 
hot spots in mini lobes
and we examine the case of $\xi_{p}=100$ in this work.
The timescale of proton synchrotron is
$t_{p,\rm syn,hs}=(6\pi m_{p}^{4}c^{3})/(\sigma_{T}m_{e}^{2}EB^{2})
\approx 
1\times 
10^{4}(E_{p}/10^{18}~{\rm eV})^{-1}
(B_{\rm hs}/0.1~{\rm G})^{-2}$.
The high energy protons escape 
from the hot spots via sideways expansions
and they are injected into the mini lobes.
When the  escape velocity from the hot spot is
$\sim 0.3~c$, the escape timescale is accordingly
$t_{\rm esc,hs}\sim 3~(R_{\rm hs}/10^{18}~{\rm cm})~{\rm years}$. 
Therefore, the typical $E_{p,\rm max}$ is obtained by the relation
$t_{p,\rm acc,hs}\approx t_{\rm esc,hs}$ and 
it is 
\begin{eqnarray}
E_{p,\rm max}\approx 
1 \times 10^{18}
\left(\frac{\xi_{p}}{100}\right)^{-1}
\left(\frac{R_{\rm hs}}{10^{18}~{\rm cm}}\right)
\left(\frac{B_{\rm hs}}{10^{-1}~{\rm G}}\right)
~{\rm eV}  .
\end{eqnarray}
%
The Larmor radius of  high energy protons with 
$E_{p}\sim 10^{18}~{\rm eV}$
is sufficiently smaller than $R_{\rm hs}$.
Hence, the conditions of proton acceleration are satisfied.
%
%

Next, high energy protons escape from the hot spot via
sideways expansions and they injected in  the mini lobes.
Then, 
they undergo various coolings in the lobes, i.e, adiabatic loss,
$p\gamma$ interaction, and proton synchrotron cooling. 
Below, we calculate these processes 
with the Monte Carlo simulation.

\section{Monte Carlo simulation}

To calculate broadband photon spectra, 
the Monte Carlo simulation has been performed 
in this work. The numerical code has been
developed by Asano et al. (2008, 2009) and references therein.
Therefore we do not repeat it here.
Details are shown in Kino and Asano (2010). 
%
Distributions of particles and photons are assumed to be
isotropic.
We include the following physical processes:
(1) photo-pion production from protons and neutrons 
($p+\gamma\rightarrow p/n+\pi^{0}/\pi^{+}$),
(2)
pion decay ($\pi^{0}\rightarrow 2\gamma$),
($\pi^{\pm}\rightarrow \mu^{\pm}+\nu$)
and muon decay
($\mu^{\pm}\rightarrow e^{\pm}+\nu$), 
(3)
photon-photon pair production 
($\gamma+\gamma\rightarrow e^{+}+e^{-}$),
(4) 
Bethe-Heitler pair production 
($p+\gamma\rightarrow p + e^{+}+e^{-}$),
(5)
synchrotron and inverse Compton processes 
of electrons/positrons,
protons, pions, muons with Klein-Nishina cross section, 
(6)
synchrotron self-absorption for electrons/positrons, and 
(7) adiabatic expansion loss.

\section{Results}

Fig.~2 shows the photon spectra for the case of
$ L_{e}=1\times 10^{45}~{\rm erg~s^{-1}}$, 
$ 1\times 10^{44}~{\rm erg~s^{-1}}$, 
$ 1\times 10^{43}~{\rm erg~s^{-1}}$, 
$ 1\times 10^{42}~{\rm erg~s^{-1}}$, and 
$ 1\times 10^{41}~{\rm erg~s^{-1}}$
where $L_{e}$ is the injection power of non-thermal electrons.
For electron Gyro-factor, we set
$\xi_{e}=\xi_{p}=1\times 10^2$.
The injection index is assumed as $s=2$ 
both for electrons and protons.
Based on VLBI observations, 
we set the following parameters as
$R=2~{\rm pc}$,
$R_{\rm hs}=0.3~{\rm pc}$,
$B=0.1~{\rm G}$,
$v_{\rm exp}=0.1~c$,
$LS=10~{\rm pc}$, and
$\gamma_{p,{\rm min}}=
\gamma_{e,{\rm min}}=10$.
As for the luminosity and temperature of 
the standard accretion disk, we assume
$L_{\rm disk}=3 \times 10^{45}~{\rm erg~s^{-1}}$, and
$kT_{\rm disk}=10~{\rm eV}$
(e.g.,  Ostorero et al. 2009).
We impose a injection power of high energy protons $L_{p}$ being
smaller than total kinetic powers of powerful
radio galaxies $\sim 10^{47-48}~{\rm erg~s^{-1}}$ 
(e.g., Ito et al. 2008). 
We assume $ L_{p}=5\times 10^{46}~{\rm erg~s^{-1}}$ 
in this work.

Thin solid lines in Fig.~2 shows
the photon spectra without proton injection (i.e., $L_{p}=0$).
The synchrotron spectra from the 
primary electrons are well studied by VLBI observations.
The turnover frequency is mainly caused by 
synchrotron self absorption (SSA)  (e.g., Snellen et al. 2000).
According to the synchrotron luminosity of the 
mini lobe $L_{\rm syn}$,
the observed turnover frequencies of mini lobes
show $\nu_{\rm SSA} \sim 0.1-10~{\rm GHz}$.
At the same time, primary electrons 
suffer from  fast synchrotron cooling.
Therefore a break frequency due to the synchrotron cooling 
typically appears comparable to or below  
$\nu_{\rm SSA}$. 
This implies that $L_{e}\approx L_{\rm syn}$.

Thick solid lines in Fig.~2 represent
the photon spectra with proton injection.
It is found that the amount of primary
electrons controls whether hadronic emission is visible or not.
For the lobe with $L_{e}=1\times 10^{45}~{\rm erg~s^{-1}}$,
the leptonic emissions overwhelm the  hadronic 
emissions at all energy domains.
As a result, the spectrum with  proton injection
is almost the same as the one without protons injection.
Therefore, the bright lobe
is not suitable for testing the existence of high energy protons.
As  $L_{e}$ 
decreases to $L_{e}\le 1\times 10^{44}~{\rm erg~s^{-1}}$,
thick and thin lines become separable at $\gamma$-ray energy 
range because synchrotron emission from
secondary electrons is added in the thick lines.
For the lobes with
$L_{e}<1\times 10^{43}~{\rm erg~s^{-1}}$,
the proton synchrotron bump appears 
in sub-MeV range with the peak at
\begin{eqnarray}
\nu_{p,\rm syn}\approx
0.7~
\left(\frac{E_{p}}{10^{18}~{\rm eV}}\right)^{2}
\left(\frac{B}{10^{-1}~{\rm G}}\right)
~{\rm MeV}.
\end{eqnarray}
%
In TeV energy range, it is shown that TeV-$\gamma$ 
photons are significantly
attenuated due to soft photons in the lobes.

\section{Summary}

Hadronic emission at extragalactic mini radio lobes is discussed
in this work. 
We calculate the broadband photon spectra taking 
$p\gamma$ interaction and proton synchrotron cooling into account.
The powers of standard accretion disk and injected protons are
$L_{\rm disk}=3\times 10^{45}~{\rm erg~s^{-1}}$,
$L_{p}=5\times 10^{46}~{\rm erg~s^{-1}}$, respectively.

For bright lobes with 
$L_{e}\sim L_{\rm syn}\sim 10^{45}~{\rm erg~s^{-1}}$, 
the predicted VHE emission is detectable with {\it Fermi}/LAT,
and HESS.
Since it is overwhelmed by the leptonic components in this case,
it is hard to figure out whether the $\gamma$-ray emission is
hadronic- or leptonic-origin. 
%
%
For the lobe with $L_{e}\sim 10^{44}~{\rm erg~s^{-1}}$, 
secondary $e^{\pm}$ pairs radiate
TeV $\gamma$ synchrotron emission
and it is detectable by the current TeV $\gamma$ telescopes.

For less luminous radio lobes
$L_{e}\sim L_{\rm syn}< 10^{44}~{\rm erg~s^{-1}}$, 
the proton synchrotron bump appears  at sub MeV energy band.
The synchrotron bump from
secondary $e^{\pm}$ pairs created by $\mu$-decay appears
at GeV/TeV ranges.
If the double bumps due to these hadronic processes
are detected, it would be a support for the hadronic model
since leptonic model does not easily account for such double bumps.

\section{Future prospects}

We add to comment on the 
recent radio observation of mini lobe 3C 84
which is associated with NGC 1275 ($z=0.0176$).
It shows the outburst around 2005 and 
the new component C3 emerges with the size 
smaller than $1~{\rm pc}$ (Nagai et al. 2010). 
{\it Fermi}/LAT also detect 
GeV $\gamma$-ray emission 
in the same period (Abdo et al. 2009).
Fig.~3 displays the actual multi-epoch images of 3C 84
obtained by VERA observations at 22 GHz (Nagai et al. 2010).
While the flux of C3 component measured by VERA increases,
GeV $\gamma$-ray flux over the period August 2008 - August 2009 
seems to remain constant (Kataoka et al. 2010).
The difference of the measured 
flux variations between VLBI and {\it Fermi}/LAT
could indicate a breakdown of the 
one-zone synchrotron-self Compton model proposed
in Abdo et al. (2009). 
Hadronic emissions could partly 
contribute in $\gamma$-ray energy bands, although we 
did not deal with the specific case of 3C 84 in this work.
In order to nail down the real $\gamma$-ray emitter,
continuous observations by VLBIs are quite important.

In the near future, the  project of 
VLBI Space Observatory Programme-2 (VSOP-2) 
with the high angular resolution (Tsuboi et al. 2009) 
will play a unique role for mini lobe observations.
Observations will be made with the 9-m antenna
in orbit together with ground radio telescopes, 
and will achieve the angular resolutions of 40, 80,
and 210~microarcsec at 43, 22, and 8~GHz, respectively.
Especially, observations at 8~GHz will unveil
fine structures of unresolved mini lobes.
Furthermore,
the era of VSOP-2 will overlap the
duration in which 
the next generation TeV $\gamma$-ray telescope {\it CTA} 
would be in operation
(http://www.cta-observatory.org/).
{\it  CTA} will have a
factor of 5-10 improvement in sensitivity in the current
energy domain of about 100 GeV to some 10 TeV
and an extension of the accessible energy range well
below 100 GeV and to above 100 TeV.
A future collaboration between VSOP-2 and {\it CTA} 
will be  one of the best ways to explore the
origin of $\gamma$-ray emission from mini radio lobes.
%


%
   \begin{figure}
   \centering
   \vspace{247pt}

   \includegraphics{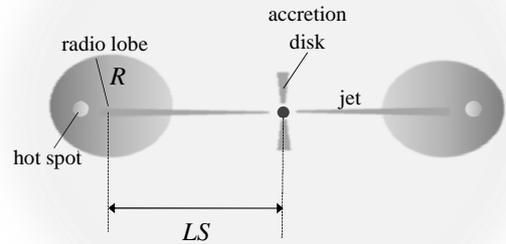}
      \caption{
         \label{fig:ptolemy}
A cartoon of mini radio lobes immersed in radiation field
due to the accretion disk.
In this work, we examine the case of 
$R= 2~{\rm pc}$,
$LS= 10~{\rm pc}$ and
$L_{\rm disk}=3\times 10^{45}~{\rm erg~s^{-1}}$.
         }
   \end{figure}
%


   \begin{figure}
   \centering
   \vspace{247pt}

   \includegraphics{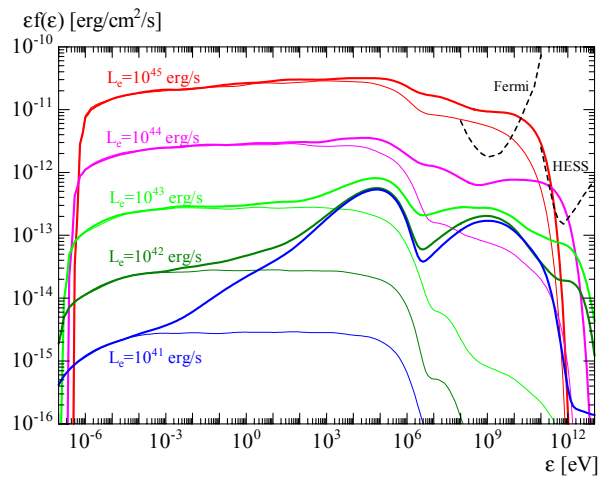}
      \caption{
Predicted broadband spectra
from the mini radio lobe for
$L_{p}=5\times 10^{46}~{\rm erg~s^{-1}}$ and
$L_{\rm uv}=3\times 10^{45}~{\rm erg~s^{-1}}$.
Black, green, yellow, red, and blue lines are
the case for
$L_{\rm e}=1\times 10^{45}~{\rm erg~s^{-1}}$, 
$1\times 10^{44}~{\rm erg~s^{-1}}$, 
$1\times 10^{43}~{\rm erg~s^{-1}}$, 
$1\times 10^{42}~{\rm erg~s^{-1}}$, and 
$1\times 10^{41}~{\rm erg~s^{-1}}$, 
respectively.
The lobes are located at the distance $D=100~{\rm Mpc}$.
The proton synchrotron bump is predicted at $\sim$ sub MeV 
for smaller $L_{e}$.
         \label{fig:ptolemy}
         }
   \end{figure}
%

   \begin{figure}
   \centering
   \vspace{247pt}
   \includegraphics{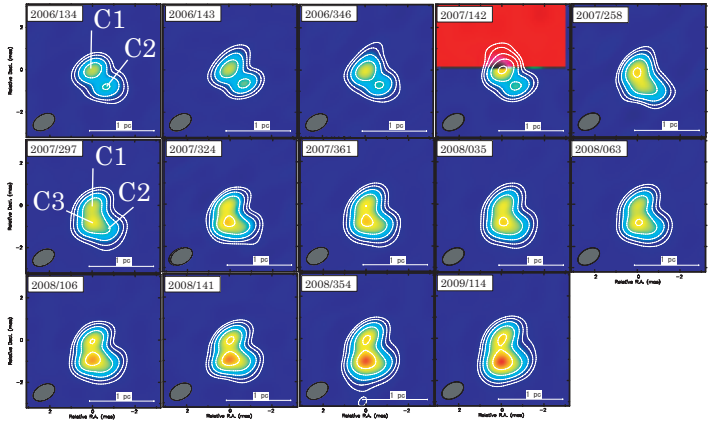}
      \caption{
Multi-epoch VERA images of 3C 84 at 22 GHz. 
The C1 component is assumed to be the core.
The C3 component is the new born one and it 
proceeds to the south. The size is still less than 1 pc.
         }
   \end{figure}
%

\begin{acknowledgements}
We are indebted to H. Nagai for providing us the image 
in Fig.~3.

\end{acknowledgements}

\end{document}